\documentclass[a4paper,fleqn]{cas-dc}
\usepackage[numbers]{natbib}

\usepackage{amsmath,amssymb,amsfonts,amsthm}
\usepackage{textcomp}
\usepackage{booktabs,tabularx}

\usepackage{float,graphicx,enumerate}

\usepackage{listings}
\usepackage{hyperref}
\usepackage[table,xcdraw]{xcolor}
\usepackage{hyphenat}
\usepackage{bbold}
\usepackage[export]{adjustbox}
\usepackage{amssymb}
\usepackage{color,soul}
\usepackage{algorithm}
\usepackage{algpseudocode}

\def\tsc#1{\csdef{#1}{\textsc{\lowercase{#1}}\xspace}}
\tsc{WGM}
\tsc{QE}
\algdef{SE}[SWITCH]{Switch}{EndSwitch}[1]{\textbf{switch} (#1)}{\textbf{end switch}}
\algdef{SE}[CASE]{Case}{EndCase}[1]{\textbf{case} #1}{\textbf{end case}}

\definecolor{unetgray}{gray}{0.95}
\newcolumntype{C}[1]{>{\centering\arraybackslash}m{#1}}

\newcommand{\unetvar}[1]{\rowcolor{unetgray}\hspace{1em}#1}

\begin{document}
\let\WriteBookmarks\relax
\def\floatpagepagefraction{1}
\def\textpagefraction{.001}

\shorttitle{Geometric Analysis of Magnetic Labyrinthine Stripe Evolution}    

\shortauthors{V.Y. Okubo et~al.}  

\title [mode = title]{Geometric Analysis of Magnetic Labyrinthine Stripe Evolution via Deep Learning Segmentation}  

\author[1]{Vinícius Yu Okubo}[
    orcid = 0009-0009-8324-3772
]
\cormark[1]
\ead{ViniciusOkubo@usp.br}
\credit{Methodology, Software, Formal analysis, Data curation, Visualization, Writing – original draft}

\author[2]{Kotaro Shimizu}
\credit{Conceptualization, Methodology, Formal analysis, Writing – original draft}

\author[3]{B. S. Shivaram}
\credit{Conceptualization, Investigation, Writing – review \& editing}

\author[3]{Gia-Wei Chern}
\credit{Conceptualization, Supervision, Writing – review \& editing}

\author[1]{Hae Yong Kim}
\credit{Conceptualization, Methodology, Supervision, Writing – review \& editing}

\affiliation[1]{organization={Dept. Electronic Systems Engineering, Polytechnic School, University of São Paulo},
            city={São Paulo},
            postcode={05508-010}, 
            state={SP},
            country={Brazil}}
\affiliation[2]{organization={Department of Applied Physics, The University of Tokyo},
            city={Tokyo},
            postcode={113-8656}, 
            country={Japan}}
\affiliation[3]{organization={Department of Physics, University of Virginia},
            city={Charlottesville},
            postcode={22904}, 
            state={Virginia},
            country={USA}}

\cortext[1]{Corresponding author}

\begin{abstract}
Labyrinthine stripe patterns are common in many physical systems, yet their lack of long-range order makes quantitative characterization challenging. We investigate the evolution of such patterns in bismuth-doped yttrium iron garnet (Bi:YIG) films subjected to a magnetic field annealing protocol. A U-Net deep learning model, trained with synthetic degradations including additive white Gaussian and Simplex noise, enables robust segmentation of experimental magneto-optical images despite noise and occlusions. Building on this segmentation, we develop a geometric analysis pipeline based on skeletonization, graph mapping, and spline fitting, which quantifies local stripe propagation through length and curvature measurements. Applying this framework to 444 images from 12 annealing protocol trials, we analyze the transition from the “quenched” state to a more parallel and coherent “annealed” state, and identify two distinct evolution modes (Type A and Type B) linked to field polarity. Our results provide a quantitative analysis of geometric and topological properties in magnetic stripe patterns and offer new insights into their local structural evolution, and establish a general tool for analyzing complex labyrinthine systems.
\end{abstract}




\maketitle

\section{Introduction}
\begin{figure}
    \centering
    \includegraphics[width=0.44\textwidth]{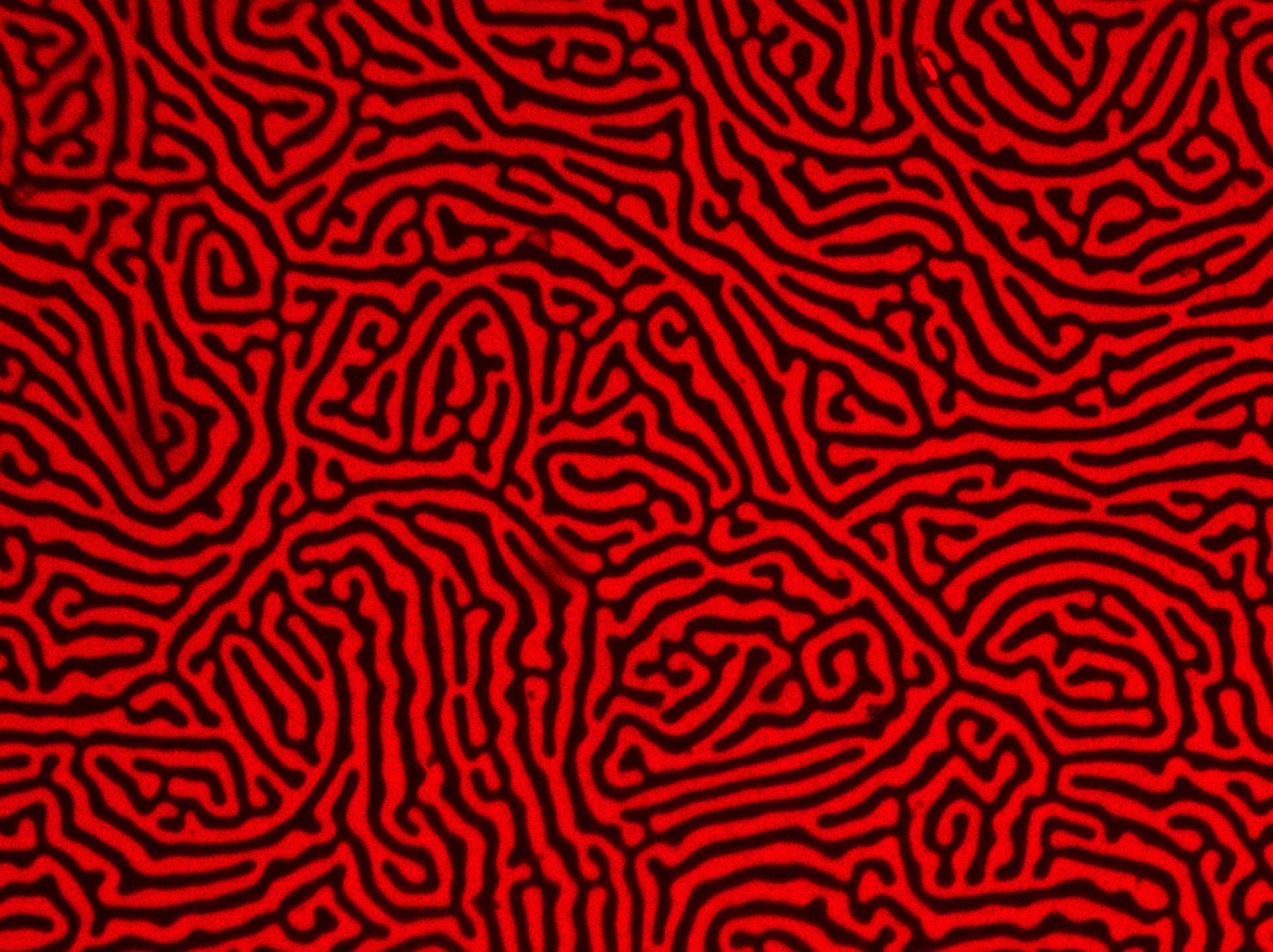}
    \caption{Magnetic stripe pattern in Bi:YIG film.}
    \label{fig:magnetic_stripe_pattern}
\end{figure}

Labyrinthine patterns, characterized by intricate stripe-like features, are hallmarks of non-linear systems that are ubiquitous in nature and appear in biological patterns, chemical reactions, crystal growth, and magnetic ordering~\cite{Nicolis77,Cross93,Koch94,Pismen06,Cross09}.
Fig.~\ref{fig:magnetic_stripe_pattern} illustrates a labyrinthine pattern in a magnetic film. 
These patterns consist of well-defined local stripes whose orientations vary randomly over large regions and are disrupted by numerous defects, resulting in the absence of long-range order.
Because of this lack of long-range correlation, defining a conventional order parameter to characterize their structure is challenging, so new methods of characterization beyond traditional phase descriptions are required.

Indeed, numerous attempts have been made to characterize labyrinthine patterns by using, e.g., structure factors (Fourier transform of labyrinthine patterns)~\cite{Ouyang91,Elder92,Cross95,Christensen98,Alar20} and disorder functions~\cite{Gunaratne95,Hu04,Nathan05,Hu05}. 
These methods utilize the global features of labyrinthine patterns and have provided simple measures of disordered structures. 
Meanwhile, they can potentially fail to capture the local structural properties such as the defects in local stripes despite their importance in disordered systems.

This difficulty hinders the understanding of the physical properties associated with such patterns, including magnetic, electric, and thermal properties, and thus a new and efficient characterization method applicable to such a disordered system is highly anticipated in material science, prompting numerous attempts thus far~\cite{Elder92,Ouyang91,Cross95,Christensen98,Gunaratne95,Hu04,Hu05,Hou97,Seul92,Seul92b,Seul92c}. 

In stripe pattern formation, defects generally exhibit unique geometrical properties associated with symmetry breaking of the local order parameter, and these properties impose strong geometrical constraints on the evolution of patterns. 
In the case of labyrinthine patterns, two different types of defects called junctions and terminals are known to play crucial roles in their formation~\cite{Seul92,Seul92b,Seul92c}. 
A single defect cannot be removed by a continuous deformation of labyrinthine patterns, while pairs of different defects can be pair annihilated due to their geometrical properties. 
Typically, physical systems evolve over time toward energetically favorable, less disordered configurations. However, due to the necessity of pair annihilation of defects for transitions to more ordered states, global structural reconfigurations become kinetically suppressed. This leads to stationary labyrinthine patterns, despite the presence of numerous defects and the absence of long-range order.

Beyond the abundance of defects in labyrinthine structures, these physical and geometrical considerations underscore the significance of analyzing local configurations within such patterns.
Since labyrinthine patterns are inherently disordered, analyzing only a small portion of the pattern is insufficient to extract meaningful features.
It is necessary to examine sufficiently large regions for unbiased analysis, and thus large-scale image analysis becomes indispensable.
The swift and systematic analyses of large-scale datasets are expected to unveil hidden features of their complicated structures, grounded in statistical analyses.

\begin{figure}
    \centering
    \includegraphics[width=0.48\textwidth]{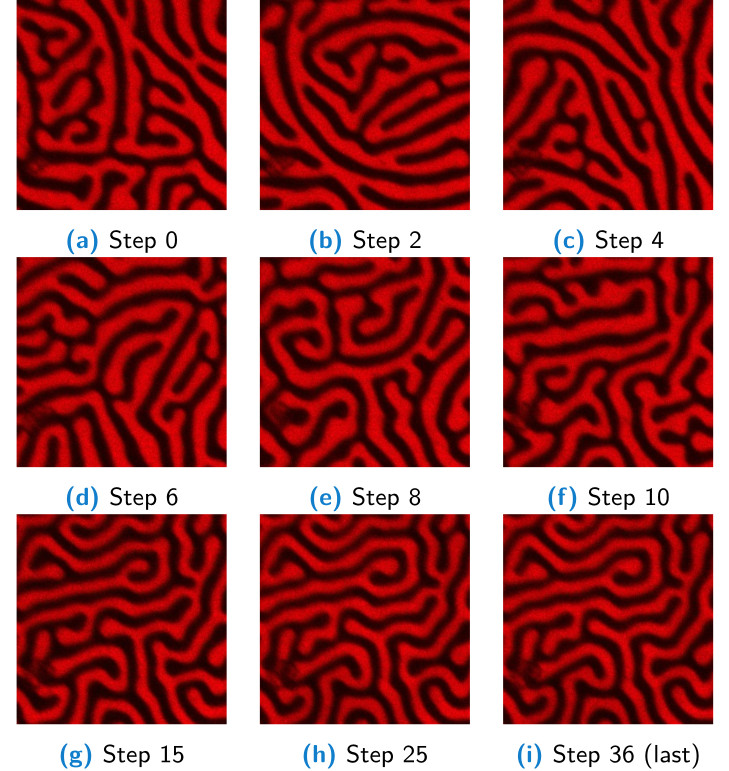}
    \caption{Magnetic field annealing transition. The crops are from the same region at different demagnetization steps.}
    \label{fig:field_annealing}
\end{figure}

Conveniently, labyrinthine patterns are seen to arise spontaneously in films of bismuth-doped yttrium iron garnet (Bi:YIG), even under zero magnetic field (Fig.~\ref{fig:magnetic_stripe_pattern}). 
The patterns represent regions of internal magnetic moments of opposing polarity, distinguished as dark and bright regions in magneto-optics experiments.
Through the application of a magnetic field annealing protocol, the stripe pattern can be made to evolve into various configurations, as shown in Fig.~\ref{fig:field_annealing}.
In the initial steps, referred to as the “quenched state,” the pattern exhibits a higher degree of spatial disorder, characterized by significant variations in stripe width and erratic propagation.
As the annealing protocol progresses, the stripe pattern moves towards a more compact and spatially coherent configuration referenced as “annealed state”.

To characterize the junctions and terminals in Bi:YIG films, Okubo et al.~\cite{okubo2024} developed TM-CNN, a technique that leverages template matching and CNN (Convolutional Neural Network) classification for accurate topological defect identification. 
Building on this technique, Shimizu et al.~\cite{shimizu2023} conducted a systematic analysis of the topological defects, demonstrating that their spatial correlations serve as an effective metric for capturing the spatial evolution of labyrinthine patterns.

\begin{figure*}
    \centering

    \includegraphics[width=0.96\textwidth]{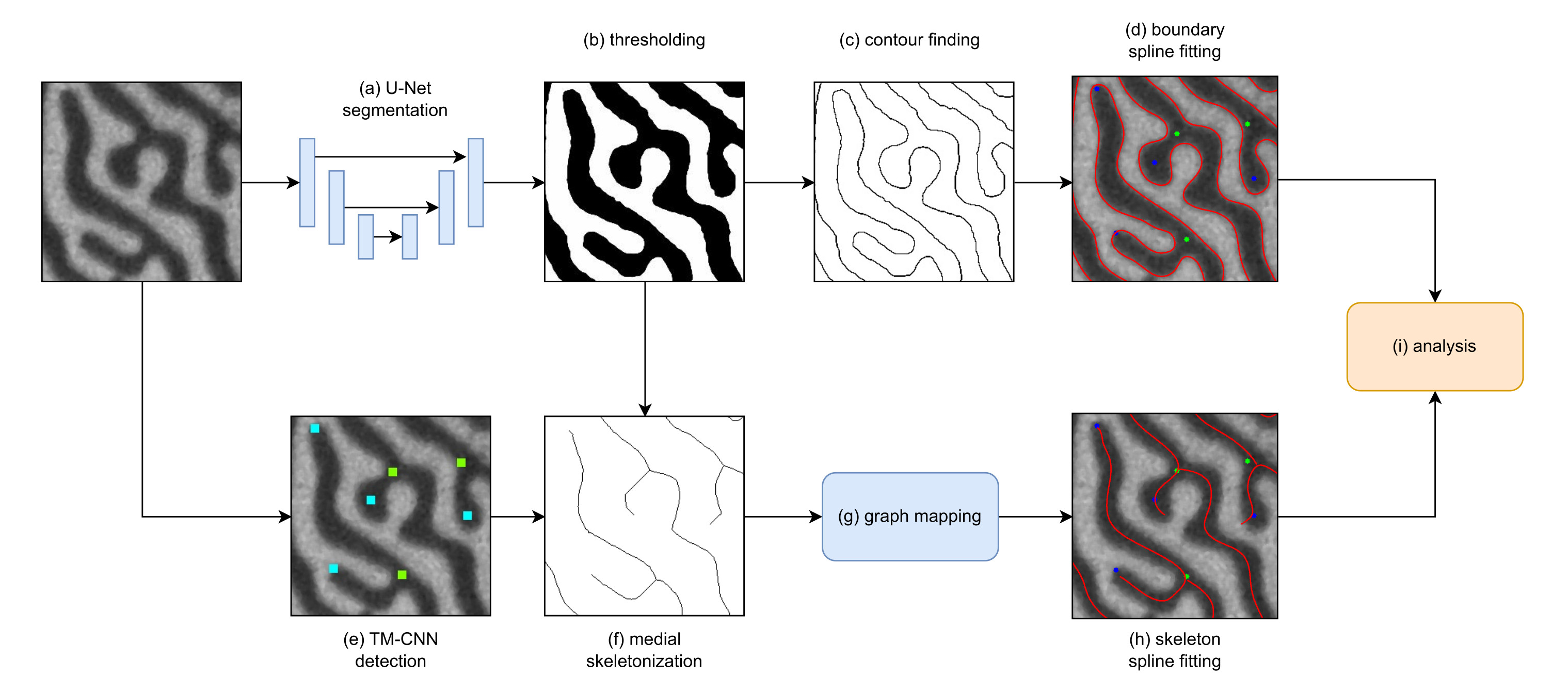}
    \caption{(a) The image is segmented with a U‑Net. (b) Threshold is applied to binarize the image. (c) The border path of the dark regions are extracted through contour finding. (d) A spline is fitted to the contour for modeling the real space contour. (e) Topological defects are detected with TM-CNN. (f) Medial skeletonization is performed to characterize the inner path along the dark regions. (g) The skeletons are mapped into a graph. (h) Spline fitting is performed for modeling the real space paths between defects. (i) The real space length of the border and inner paths are used to characterize the magnetic labyrinthine patterns.}
    \label{fig:image_pipeline}
\end{figure*}

While previous studies emphasized the topological characterization of labyrinthine patterns, this work proposes a new framework for analyzing the geometric features of stripe propagation, with particular attention to the segments between topological defects. The analysis was performed in a large scale dataset containing 444 images of Bi:YIG films from 12 separate field annealing protocols. 
Specifically, we evaluate segment length and curvature to describe how stripes evolve within the labyrinthine pattern. 
This geometric perspective not only highlights the physical mechanisms underlying stripe formation but also offers insights into the control of the structure and resultant material properties.

Identifying stripe propagation requires precise segmentation between dark and light areas. 
In real-world scenarios, this segmentation is complicated by image degradations such as noise, blurred regions, and occlusions, which conceal important details in the formation of the stripes.
We propose a modern deep-learning filter approach to perform robust segmentation of the magnetic stripe pattern.

We combine image restoration (IR) and segmentation techniques to train a model on synthetically degraded images, achieving accurate segmentation of the stripe pattern formed on Bi:YIG films. 
While degradations such as blur, rain, and Gaussian noise are commonly used within image restoration, we demonstrate that Simplex noise~\cite{perlin2002} can simulate smudge-like occlusions.
The resulting segmentations were mapped into graph representations of the magnetic pattern, efficiently storing topological defect locations along with the length and curvature of the stripe segments connecting them.
Through geometric analysis, we characterize the propagation of magnetic stripes, uncovering new aspects of their geometric and topological organization.
In particular, we characterize an interleaved behavior resulting from the field annealing protocol, find particular geometric properties in the quenched to annealed transition, and identify geometric differences between junction-junction and junction-terminal stripe segments.

\section{Related works}

Image segmentation is essentially classifying each pixel in an image.
Long et al. \cite{long2015} proposed Fully Convolutional Networks, a fully convolutional architecture with end-to-end training for image segmentation.
It employs an encoder-decoder architecture, sequentially generating higher-level features from the input image until applying segmentation.
The U-Net model proposed by Ronneberger et al.~\cite{ronneberger2015} introduced a U-shaped encoder-decoder CNN architecture for image segmentation.
Through the usage of concatenation bridges, the network can leverage jointly detailed spatial information from low-level features and semantic context from high-level features.
This structure enables highly sample-efficient training while maintaining a low memory footprint relative to the number of parameters.

Since its introduction, U-Net has become a widely used architecture across various tasks.
It is particularly popular for medical image segmentation~\cite{azad2024} and has also demonstrated success in general microscopy image segmentation~\cite{Wu2022}.
Still, most research on image segmentation has focused on clean and well-defined images.
Some studies have addressed various types of degradations, such as white noise, blurring, haze, rain, and low-light conditions~\cite{xia2019, guo2019, rajagopalan2023}.
However, image segmentation in settings with visual obstructions or occlusions remains relatively underexplored.

Degradations are more commonly tackled in the field of image restoration, which is dedicated in obtaining high-quality, clear images from degraded ones.
Most deep learning approaches rely on synthetic degradations to create clean and degraded image pairs for model training.
These degradations do not necessarily reflect real-world conditions, so creating more representative degradations remains a challenge, motivating the development of many domain-specific approaches~\cite{zhao2017, matsui2019, sakaridis2018, hao2019}.

Additive White Gaussian Noise (AWGN) filtering is a major area of interest in image restoration.
Early neural network-based approaches, such as Multi-Layer Perceptrons (MLPs), showed promise in noise filtering tasks by matching the performance of previous techniques at a single noise level, yet they struggled with variable noise levels \cite{burger2012}.
Since then, CNNs have not only become widely adopted for AWGN filtering, but have also proved highly successful.
Zhang et al.~\cite{zhang2017} introduced DnCNN, a CNN model designed for Gaussian noise removal through residual image extraction.
Later, Zhang et al.~\cite{zhang2018} proposed FFDNet, which further improved Gaussian noise filtering by incorporating noise level maps as inputs, enabling better adaptability to varying noise levels across an image.
Park et al.~\cite{park2019} introduced a U-Net-based model for noise filtering.
To handle different noise levels, their model was trained using Gaussian noise augmentation with varying variance settings.

In contrast, Simplex and Simplex-like noises have been used more sporadically in task-specific settings.
Perlin noise \cite{perlin1985}, which is similar to simplex noise, was proposed for generating random Region of Interest patches in HRCT images \cite{Bae2018}.
In chest X-ray classification of lung cancer, Perlin noise has also been combined with AWGN as an augmentation strategy \cite{Haekal2021}.
Beyond medical imaging, Perlin noise has been adopted to simulate haze for dehazing tasks \cite{Zheng2024}.
Mahmood et al.~\cite{Mahmood2024} applied simplex noise to define the shape of synthetic cracks for segmentation of structural cracks.
More recently, it has been employed to generate anomalies for industrial image anomaly detection \cite{Li2025}.

Transformer-based models demonstrate superior generalization capabilities compared to CNNs when trained with sufficient data and computational resources.
Transformers allow global interactions between input tokens through the attention mechanism, while convolution-based approaches are biased toward local patterns due to their translation-invariant receptive fields.
Vision Transformers (ViT) were introduced by Dosovitskiy et al.~\cite{dosovitskiy2020} for image classification, representing images as sequences of patch embeddings processed by a pure transformer encoder.
Building on ViT, several transformer-based architectures have been proposed for segmentation.
SETR~\cite{zheng2021} uses a ViT encoder with convolutional decoders for semantic segmentation, while TransUNet~\cite{chen2021Transunet} combines a CNN encoder with transformer blocks and a U-Net-style decoder.
Segmenter~\cite{strudel2021} employs a ViT backbone and a transformer-based decoder to produce class masks.
Swin Transformer~\cite{liu2021} introduces a hierarchical windowed attention backbone, which has been adapted to a U-shaped design in Swin-Unet~\cite{cao2022} for segmentation.
SegFormer~\cite{xie2021segformer} is a hierarchical transformer-based model whose encoder produces multiscale features using overlapping patch embeddings, while a simple all-MLP decoder upsamples and fuses features into high-resolution segmentation maps without positional encodings.

Alternative techniques for IR have also been proposed by leveraging the transformer architecture ~\cite{chen2021,liang2021,zamir2022} and generative models.
Isola et al.\cite{isola2017} introduced pix2pix, a framework based on Generative Adversarial Networks (GANs)~\cite{goodfellow2014} for general image-to-image translation.
Their flexible approach allows for its usage in both image restoration and image segmentation tasks.
Alternatively, Majurski et al. used trained GAN discriminators for pre-optimizing U-Nets for segmenting cell images\cite{majurski2019}.
Diffusion models have also been studied for image restoration\cite{kawar2022, luo2023, xie2023}.

Deocclusion is a subfield within image restoration commonly associated with inpainting. 
Many single-image deocclusion approaches rely on a two-step procedure: first identifying occlusion locations, then inpainting the affected areas.
For example, Matsui and Ikehara~\cite{matsui2019} utilized a U-Net model to segment fence-occluded regions.
For raindrop removal, Qian et al.~\cite{qian2018} introduced a GAN-based method that generates a heatmap of potential raindrop locations and subsequently produces an image without raindrops using conditional generation.
Dong et al.\cite{dong2020} also adopted a two-step framework for face deocclusion, combining U-Nets with GAN training for occlusion segmentation and inpainting.
Other approaches perform deocclusion in an unified model.
Angah and Chen~\cite{angah2020} approached construction site deocclusion by directly training a single U-Net model using GAN framework for training.
In the case of face deocclusion, Zhao et al.~\cite{zhao2017} proposed a model that uses an LSTM-Autoencoder trained on paired clean and synthetically occluded images, jointly learning to detect and restore occlusions.

\section{Methodology}

To analyze the magnetic labyrinthine pattern, we developed a processing pipeline illustrated in Fig.~\ref{fig:image_pipeline}.
Segmentation of the light and dark regions is performed through a U-Net model.

The U-Net model was chosen for its widespread use in microscopy segmentation~\cite{Wu2022}.
Although Transformer-based approaches have recently gained popularity due to their ability to model global context, the propagation of labyrinthine patterns is primarily governed by their local structure.
Similarly, noise and occlusion degradations in labyrinthine images are spatially confined.
U-Net’s local inductive bias is thus a good match for this task, supporting strong generalization in this data-constrained setting.

Two distinct types of paths are then characterized: 
\begin{enumerate}
\item Border paths, which follow the boundary between light and dark regions (Fig. \ref{fig:image_pipeline}(d)). 
\item Inner paths, which traverse the center of the dark regions, divided by the pattern's junctions and extending until its terminals (Fig. \ref{fig:image_pipeline}(h)).
\end{enumerate}

Spline fitting is used to generate smooth curves defined in $\mathbb{R}^2$ space following the extracted paths.
Measurements of the length and curvature are then processed to characterize properties of the magnetic labyrinthine pattern.

\subsection{Dataset}

Images of labyrinthine magnetic domains were acquired from thin films of bismuth‑doped yttrium iron garnet subjected to a controlled magnetic‑field annealing protocol.
Each annealing trial began by saturating the film’s magnetization with a strong external out‑of‑plane field ($\pm z$).
The field was then switched off and held at zero for $10$ s, allowing the magnetization to relax into a domain structure with opposing polarities.
The resulting domain pattern was captured using polarized‑light microscopy\cite{Puchalska78}.

\begin{figure}
    \centering
    \includegraphics[width=0.48\textwidth]{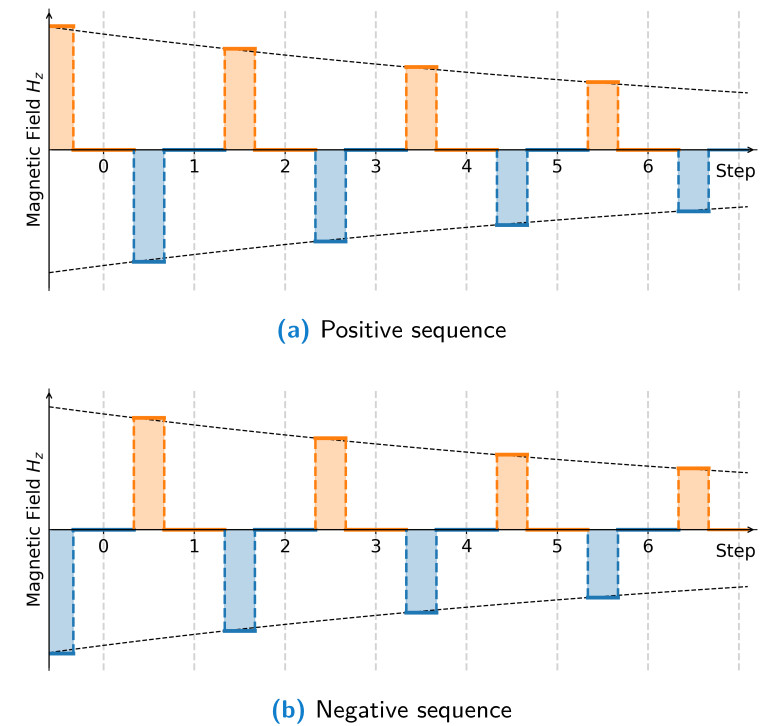}
    \caption{Field annealing protocol for positive and negative sequences. Images “Type~A” are obtained after applying upward-pointing magnetic field (orange) and images “Type~B” are obtained after applying downward-pointing magnetic field (blue).}
    \label{fig:process}
\end{figure}

We define a step as applying the external field, setting it to zero, and taking a photograph.
Each trial consists of 37 steps: the field amplitude decays exponentially from step to step, and its polarity alternates.
Two trial types were run.
In the positive sequence, the initial saturating field pointed along $+z$, whereas in the negative sequence it pointed along $-z$ (Fig.~\ref{fig:process}).

Images taken after a $+z$ field are labeled Type~A (orange).
Those taken after a $-z$ field are labeled Type~B (blue).
Because the applied field alternates directions every step, Type~A and Type~B images are interleaved.
In the positive sequence, Type~A images correspond to the even-numbered steps (0, 2, 4, …) and Type~B images correspond to the odd-numbered steps (1, 3, 5, …).
This assignment is reversed in the negative sequence: Type~A images correspond to odd-numbered steps, and Type~B images correspond to even-numbered steps.

Each sequence type was repeated six times, yielding 12 complete trials and a total of 444 domain images with $5200{\times}3888$ resolution.
Images can then be indexed by the tuple:
\begin{equation}
I \equiv (t, \sigma,s)
\end{equation}
$t\in\{1,\dots,6\}$ denotes the trial, $\sigma\in\{+,-\}$ is the sequence type (positive/negative), and $s\in\{0,\dots,36\}$ is the magnetic field annealing step index.
All measurements were performed at room temperature on a $2\text{mm}{\times}1.8\text{mm}$ film.
Table~\ref{tab:dataset_overview} summarizes the dataset characteristics.

\begin{table}[h]
    \centering
    \caption{Dataset overview.}
    \label{tab:dataset_overview}
    \begin{tabularx}{\columnwidth}{l X}
        \toprule
        \multicolumn{2}{l}{\textbf{Index Parametrization}} \\[-.5ex]
        \midrule
        Trial index ($t$) & $\{1,2,3,4,5,6\}$\\
        Sequence ($\sigma$) & $\{+,-\}$ \\
        Step index ($s$) & $\{0,1,2,\dots,36\}$ \\
        \midrule
        \multicolumn{2}{l}{\textbf{Types in Positive Sequence ($\sigma \in \{+\}$)}, indexes $s$} \\[-.5ex]
        \midrule
        Type A & $\{0,2,4,\dots,36\}$ \\
        Type B  & $\{1,3,5,\dots,35\}$ \\
        \midrule
        \multicolumn{2}{l}{\textbf{Types in Negative Sequence ($\sigma \in \{-\})$}, indexes $s$} \\[-.5ex]
        \midrule
        Type A & $\{1,3,5,\dots,35\}$ \\
        Type B  & $\{0,2,4,\dots,36\}$ \\
        \midrule
        \multicolumn{2}{l}{\textbf{Image properties}} \\[-.5ex]
        \midrule
        Resolution & $5200\text{px} {\times} 3888 \text{px}$ \\
        Film size & $2\text{mm} {\times} 1.8\text{mm}$\\
        Total images & $(6+6)\times37=444$ \\
        \bottomrule
    \end{tabularx}
\end{table}

\subsection{Segmentation}

Prior to U-Net-based segmentation, we attempted a simple local Otsu thresholding algorithm, which determines local thresholds by maximizing the inter-class variance between intensity domains \cite{otsu1975}.
Table~\ref{tab:segmentation_examples} shows a side-by-side comparison of the Otsu and U-Net segmentations on both clear and degraded patches.

In magnetic stripe patterns, regions with high local intensity variance exhibit a greater intensity separation between domains and a clearer bimodal intensity distribution
(Table~\ref{tab:segmentation_examples}, first row).
As a result, segmentation in such regions is generally consistent with the visually identifiable stripe morphology.

However, in regions with reduced illumination, brightness and contrast decrease, making the intensity distributions of magnetic domains less separable (Table~\ref{tab:segmentation_examples}, second row).
This reduced contrast amplifies the relative impact of sensor noise, hindering threshold-based segmentation.
Additionally, structured occlusions partially mask the stripe pattern, introducing localized distortions in the segmentation  (Table~\ref{tab:segmentation_examples}, third row).
Although occlusions may appear throughout the image, in practice they tend to introduce textures with intermediate intensities, reducing the local variance as well.

The U-Net model was then developed to perform robust segmentation under these degradations.
Training was performed using high-variance regions of the image with stronger domain separability, and the corresponding Otsu segmentations were used as pseudo-labels (“ground-truth proxies”).
Synthetic degradations were then applied to the input images while keeping the pseudo-labels fixed, so that the network could learn to recover the stripe structure under noise and occlusion without inheriting the limitations of threshold-based segmentation.

\begin{table}
    \caption{Comparison of segmentation methods across clear, noisy, and occluded regions. U-Net outperforms Otsu thresholding in settings under noise and occlusions.}
    \label{tab:segmentation_examples}
    \centering
    \begin{tabularx}{\columnwidth}{
        @{}
        C{0.12\columnwidth}
        C{0.24\columnwidth}
        C{0.24\columnwidth}
        C{0.24\columnwidth}
        @{}
    }
        \toprule
        Condition & Original & Otsu & U-Net \\
        \midrule
        Clear
            & \includegraphics[width=\linewidth,valign=c]{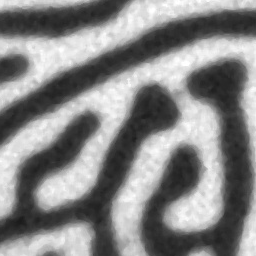}
            & \includegraphics[width=\linewidth,valign=c]{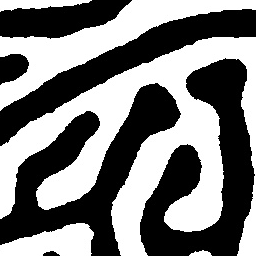}
            & \includegraphics[width=\linewidth,valign=c]{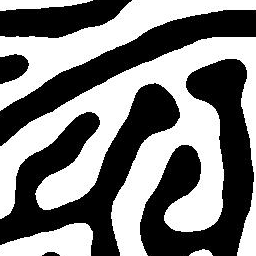} \\
        \midrule
        Noisy
            & \includegraphics[width=\linewidth,valign=c]{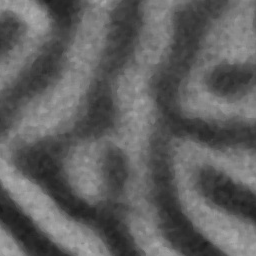}
            & \includegraphics[width=\linewidth,valign=c]{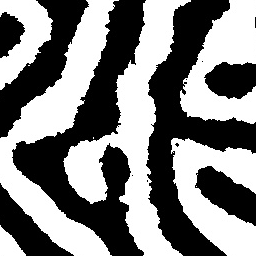}
            & \includegraphics[width=\linewidth,valign=c]{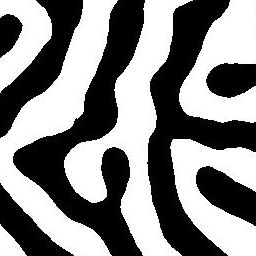} \\
        \midrule
        Occluded
            & \includegraphics[width=\linewidth,valign=c]{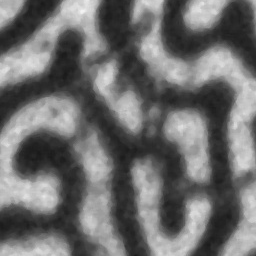}
            & \includegraphics[width=\linewidth,valign=c]{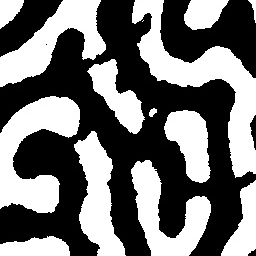}
            & \includegraphics[width=\linewidth,valign=c]{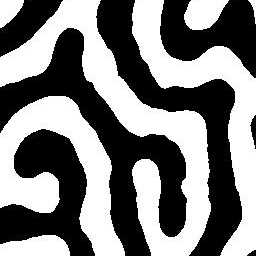} \\
        \bottomrule
    \end{tabularx}
\end{table}

\subsubsection{Pre-processing}

To train the model, we first built a dataset of image pairs containing magnetic-stripe patches and their corresponding segmentations.
These patches were sourced from the 444 experimental images.
As a pre-processing step, the images were converted to grayscale using the red channel.

\begin{figure}
    \centering
    \includegraphics[width=0.48\textwidth]{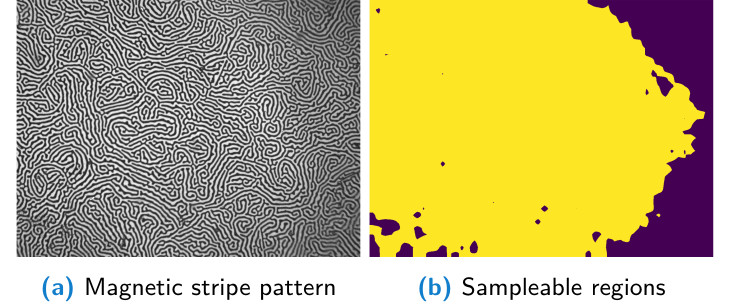}
    \caption{Regions with low noise and distinct patterns (yellow) extracted from the magnetic stripe image for U-Net model training. Purple areas represent high-noise regions that were excluded from the training dataset.
    }
    \label{fig:mask}
\end{figure}

We then extracted $320{\times}320$ patches from the high-contrast regions.
To locate these areas systematically, we computed local brightness variance with a sliding window of $64{\times}64$ pixels, advancing in 32-pixel steps.
The resulting low-resolution variance map was upscaled to the original $5200{\times}3888$ resolution with bilinear interpolation.
We sampled patches only where the local variance was above $80\%$ of the image-wide variance.
Fig.~\ref{fig:mask} shows the outcome: yellow marks the low-noise regions used for training, while purple highlights the high-noise areas that were excluded.

In order to make the model capable of segmenting the degraded regions of the magnetic stripe pattern, we synthetically introduced degradations to the clear patches used for the training.
To account for uneven illumination and contrast variations in the acquired images, we included photometric transformation augmentation consisting of brightness and gamma perturbations.
AWGN was then used to model noise degradation, as a 
simple representation of uncorrelated sensor noise commonly assumed in imaging systems \cite{gonzalez2018}.

In contrast, occlusions must be smooth, spatially correlated, and textured to resemble real smudge-like artifacts.
To generate such patterns procedurally, Perlin noise, widely used in texture generation within computer-generated imagery \cite{ebert2003}, provides a means to produce coherent, natural-appearing textures \cite{perlin1985}.
In practice, we employed Simplex noise \cite{perlin2002}, a derivative of the Perlin algorithm, to simulate occlusions due to its lower computational cost and reduced directionality artifacts.

\begin{figure}
    \centering
    \includegraphics[width=0.48\textwidth]{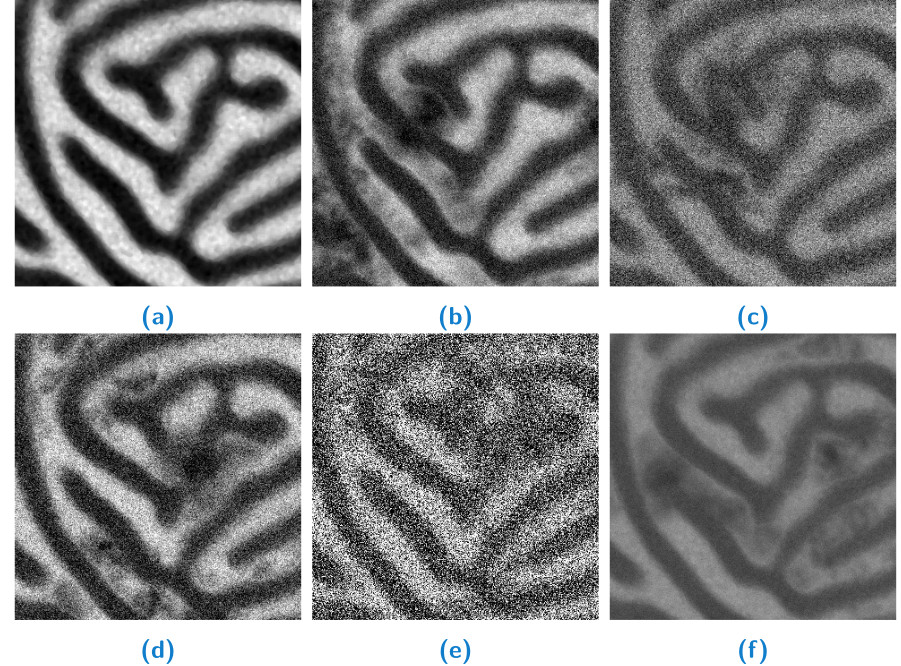}
    \caption{
    (a) Original magnetic stripe patch. (b–f) Instances of the same patch after adding various configurations of synthetic noise.
    }
    \label{fig:noise}
\end{figure}

To represent the diverse noise intensities and structural distortions found in the original images, the noise parameters were randomly sampled as an augmentation step during training (Fig.~\ref{fig:noise}).
This ensured that the network learned to segment images with different noise characteristics.

Appendix~\ref{sec:appendix_A} presents additional details on the synthetic degradations and includes a comparative table illustrating their impact on segmentation accuracy.

\subsubsection{Architecture}

The U-Net architecture employed in this study comprises two main components: an encoder and a decoder. 
The encoder systematically extracts increasingly abstract feature representations from the input image through a series of convolutional layers and max-pooling operations. 
This process progressively reduces the spatial dimensions of the feature maps while simultaneously increasing their depth, thereby capturing hierarchical information from the image.
The decoding process mirrors the encoder in reverse to generate the segmented image.
It progressively upsamples the feature maps to increase their spatial dimensions while decreasing their depth.
At each stage of decoding, we incorporate the corresponding embeddings from each step of the encoding process through skip connections.
This combines high-resolution information from the encoder with the decoder's abstracted features.

In the final step, a $1{\times}1$ convolutional layer reduces the depth of the embeddings to a single final dimension.
During inference, we apply thresholding directly after the $1{\times}1$ convolutional layer logits with $threshold=0$.

A detailed description of the model architecture is provided in Table~\ref{table:model_arch}.

\begin{table*}[ht]
\caption{Adapted U-Net architecture.}
\label{table:model_arch}
    \centering
    \begin{tabularx}{\textwidth}{
        >{\centering\arraybackslash}X 
        >{\centering\arraybackslash}c 
        >{\centering\arraybackslash}X 
    }
    \toprule
    Encoder ($\downarrow$) & Bridge & Decoder ($\uparrow$) \\
    \midrule
    Conv2D(64) + Batch Norm + ReLU  & $\rightarrow$ & Conv2D(64) + Batch Norm + ReLU + Conv2D(1)\\
    $\downdownarrows$ + Conv2D(128) + Batch Norm + ReLU & $\rightarrow$ & Conv2D(128) + Batch Norm + ReLU + $\upuparrows$ \\
    $\downdownarrows$ + Conv2D(256) + Batch Norm + ReLU & $\rightarrow$ & Conv2D(256) + Batch Norm + ReLU + $\upuparrows$ \\
    $\downdownarrows$ + Conv2D(512) + Batch Norm + ReLU & $\rightarrow$ & Conv2D(512) + Batch Norm + ReLU + $\upuparrows$ \\
    \multicolumn{3}{c}{$\downdownarrows$ + Conv2D(1024) + Batch Norm + ReLU + $\upuparrows$} \\
    \bottomrule
    \multicolumn{3}{l}{\small The encoder processes input from top to bottom ($\downarrow$), while the decoder reconstructs the output from bottom to top ($\uparrow$).} \\
    \multicolumn{3}{l}{$\rightarrow$: Concat, $\downdownarrows$: MaxPooling2D(2), $\upuparrows$: Upsampling ($\times 2$ resolution)}
    \end{tabularx}
\end{table*}

\subsubsection{Training}

The U-Net was trained to reconstruct Otsu-thresholded binary masks from clear patches with $320{\times}320$ resolution.
These patches were corrupted with synthetic noise to simulate degraded regions.
Trials $t \in \{1, 2, 3, 4\}$, comprising 296 images, were used for training and validation. 
The remaining 148 images from trials $t \in \{5, 6\}$ were reserved for the test set.

The training set comprised patches from high-local-variance regions, while validation used areas with a local variance under 70\% of the overall image variance, ensuring that most patches exhibited substantial degradation. 
We sampled and manually annotated 64 patches ($320{\times}320$).
Performance metrics were calculated without augmentation.

For each training epoch, four patches were randomly sampled from the well-defined region of each training image.
The patches were further augmented by applying random rotations of $0^\circ$, $90^\circ$, $180^\circ$ or $270^\circ$.
The network was implemented in Python using the PyTorch framework\footnote{https://pytorch.org/}.
It was trained with a batch size of 32 and a learning rate of $10^{-3}$.
The loss function was Binary Cross Entropy with Logits Loss (\texttt{BCEWithLogitsLoss}), which internally applies a sigmoid function, allowing for training with the segmentation targets in the $[0, 1]$ range.
The model was trained for a total of 30 epochs with the Adam optimizer.
The epoch with the best validation IoU was used for test evaluation.

The resulting loss curves and validation performance metrics are compiled in
Appendix~\ref{sec:appendix_B},
alongside a quantitative ablation study on the impact of synthetic degradations during training on the model's performance.

Training was conducted using a single NVIDIA A100 GPU in the Google Colab environment\footnote{colab.research.google.com}.

\subsubsection{Testing}
The U-Net’s performance was tested for segmenting degraded regions using the held-out testing trials $t \in \{5,6\}$.
Analogous to the validation procedure, 64 patches of $320{\times}320$ pixels were randomly sampled from the low-variance areas of the images 
and manually segmented to serve as the ground truth for evaluation.

Given the recent popularity in Transformer-based segmentation models, we also finetuned a SegFormer-B3 model initialized from ImageNet-1k weights.
The model was trained on the same dataset and noise augmentations as the U-Net model.

For performance evaluation, we report the F1 score, Intersection over Union (IoU), 
Average Symmetric Surface Distance (ASSD) in pixels,
and the percentage deviation in inner length ($\Delta$Inner) and border length ($\Delta$Border).
The results are summarized in Table~\ref{tab:test}.

In our testing, both the U-Net and the SegFormer-B3 model outperformed the Otsu baseline, achieving similar results in ASSD, F1-Score and IoU. 
U-Net achieved a slightly better $\Delta$Inner, while SegFormer-B3 achieved a better $\Delta$Border.
We used U-Net, as its faster processing and lower resource requirements for high resolution segmentation provides a clear practical advantage without a meaningful sacrifice in performance.

\begin{table}
    \centering
    \caption{Segmentation performance metrics.}
    \label{tab:test}
    \setlength{\tabcolsep}{3pt} 
    \begin{tabularx}{\columnwidth}{@{} l *{5}{>{\centering\arraybackslash}X} @{}}
        \toprule
        Model &
        \makebox[0pt][c]{F1 ($\uparrow$)} &
        \makebox[0pt][c]{IoU ($\uparrow$)} &
        \makebox[0pt][c]{ASSD ($\downarrow$)} &
        \makebox[0pt][c]{$\Delta$Inner} &
        \makebox[0pt][c]{$\Delta$Border} \\
        \midrule
        Otsu & 0.948 & 0.901 & 1.08 & 1.38\% & 2.90\% \\
        SegFormer-B3 & \textbf{0.953} & \textbf{0.911} & \textbf{1.01} & -0.41\% & \textbf{-0.05\%} \\
        U-Net & \textbf{0.953} & 0.910 & \textbf{1.01} & \textbf{-0.15\%} & 1.16\% \\
        \bottomrule
    \end{tabularx}
\end{table}

\subsection{ Border and inner paths }

The segmented images were processed further to extract the border and the inner paths of the dark regions.
Specifically, we generated new images containing the contour and the topological skeleton respectively.

The topological skeleton, also known as the medial axis, was generated using the \textit{skimage} implementation of medial axis skeletonization.
It represents the set of points equidistant to at least two points on the boundary using the Euclidean distance transform.
Within the magnetic stripes, it forms the central axis along the dark patterns, but includes spurious branches caused by minor variations in the border shape.
To accurately model the inner paths within the dark patterns, these spurious branches were removed.
This was done by matching the terminal points of the skeleton with the terminals detected by TM-CNN defect detection~\cite{okubo2024}.
Terminals that did not match were trimmed back to a junction point, provided their length did not exceed 45 pixels.

\begin{figure}
    \centering
    \includegraphics[width=0.48\textwidth]{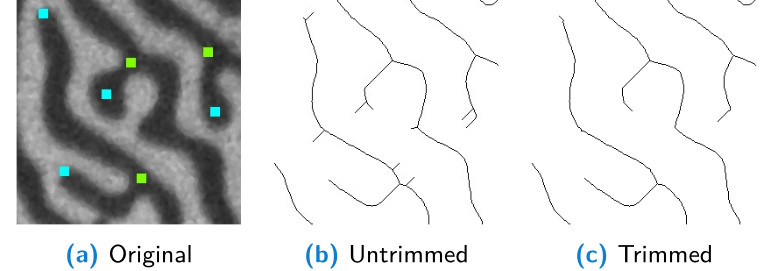}
    \caption{Topological skeleton of magnetic labyrinthine patterns.}
    \label{fig:topological_skeleton}
\end{figure}

Fig.~\ref{fig:topological_skeleton} illustrates the process:
(a) shows the original magnetic stripe pattern with TM-CNN-detected defects highlighted; (b) shows the untrimmed skeleton; and (c) shows the result after trimming.

The boundary of the stripe patterns were produced through contour finding using \textit{OpenCV}'s \textit{findContours}, which is based on the Suzuki-Abe algorithm \cite{suzuki1985}.
Fig.~\ref{fig:image_pipeline}(c, f) showcases the boundary and skeleton images respectively.

\subsection{Graph mapping}

The stripe pattern is represented as a graph where nodes denote topological defects (terminals and junctions) and edges correspond to the stripe segments connecting them. 
Each edge stores its segment's path, providing an intuitive structure for subsequent measurements.

Paths are extracted from the trimmed skeleton.
Initially, terminal pixels are found by convolving the skeleton with a $3{\times}3$ kernel whose centre is 10 and neighbors are 1.
A terminal pixel produces a convolution sum of 11.
These pixels become terminal nodes.

To process the skeleton, terminal coordinates are initially pushed onto a stack. 
We then use each coordinate as a starting point (kernel) to trace the trimmed skeleton using 8-connectivity on a temporary image copy. 
Pixels are marked as 255 upon visitation to avoid redundant processing. 
If a trace reaches another defect, we verify its representation as a graph node and create an edge connecting it to the starting kernel node, saving the path taken. 
When a junction is encountered —a pixel with several unvisited neighbors— a junction is created.
Because junctions in the pixelated skeleton consist of pixel clusters rather than single points, a node is initialized for each constituent pixel within the junction.
These nodes are connected to one another through edges with empty paths to represent a single topological structure.
Only the edges connecting the junction to outgoing branches store the corresponding segment path.
In this way, the nodes represent a single junction, preserving branch connectivity while avoiding ambiguity during path tracing.
The unvisited neighbors are then pushed onto the stack so that every branch can be traced.

If a trace ends with no neighbor in the working copy, we consult the original skeleton to disambiguate: with one original neighbor, the endpoint is a true terminal and we attach the edge to it; with two or more, the endpoint is a previously reached junction, which was reached due to a closed loop in the stripe pattern. 
In such case, we attach the edge to the hidden junction node.

The full procedure appears as pseudocode in Alg.~\ref{alg:graph_mapping}.

\begin{algorithm*}
\caption{ Graph mapping, yielding a graph representation where nodes correspond to topological defects and edges correspond to stripe segments.}
\label{alg:graph_mapping}
\begin{algorithmic}[1]
    \State $available\_path \gets copy(skeleton\_image)$
    \State $defect\_coords \gets pile(kernel\_coords)$
    \While{\Call{NotEmpty}{$defect\_coords$}}
        \State $kernel\_coord \gets defect\_coords.pop()$
        \State $path \gets [kernel\_coord]$
        \ForAll{$current\_pixel$ in $path$} \Comment{Traverse the skeleton}
            \State $available\_path[current\_pixel] \gets 255$ \Comment{Set pixel as traversed}
            \State $adjacent\_pixels \gets$ \Call{FindAdjacentPixels}{$available\_path$, $current\_pixel$}
            
            \Switch{\Call{Length}{$adjacent\_pixels$}}
                \Case{$1$} \Comment{Path continues}
                    \State $path.push(adjacent\_pixels[0])$
                \EndCase
                \Case{$>1$} \Comment{Junction in \textit{available\_path}}
                    \State \Call{AddNode}{$current\_pixel$}
                    \If{$kernel\_coord \ne $ $current\_pixel$}
                        \State \Call{AddEdge}{$kernel\_coord$, $current\_pixel$, $path$}
                    \EndIf
                    \ForAll{$adj\_pixel$ in $adjacent\_pixels$}
                        \State \Call{AddNode}{$adj\_pixel$}
                        \State \Call{AddEdge}{$current\_pixel$, $adj\_pixel$, $[current\_pixel, adj\_pixel]$}
                        \State $available\_path[adj\_pixel] \gets 255$
                        \State $defect\_coords.push(adj\_pixel)$
                    \EndFor                
                \EndCase
                \Case{$0$} \Comment{No neighbor left in \textit{available\_path}}
                    \If{$current\_pixel \ne kernel\_coord$}
                        \State $skeleton\_image\_adj\_pixels \gets$ \Call{FindAdjacentPixels}{$skeleton\_image$, $current\_pixel$}
                        \If{\Call{Length}{$skeleton\_image\_adj\_pixels$}$=1$} \Comment{True terminal in original skeleton}
                            \State \Call{AddEdge}{$kernel\_coord$, $current\_pixel$, $path$}
                        \ElsIf{\Call{Length}{$skeleton\_image\_adj\_pixels$}$\ge 2$} \Comment{Previously reached junction}
                            \State $hidden \gets$ first $q \in skeleton\_image\_adj\_pixels$ such that $q \notin path$
                            \State \Call{AddEdge}{$kernel\_coord$, $hidden$, $path$}
                        \EndIf
                    \EndIf
                \EndCase
            \EndSwitch
        \EndFor
    \EndWhile
\end{algorithmic}
\end{algorithm*}

\subsection{Length measurements}
\label{subsection:length_measurement}

\begin{figure}
    \centering
    \includegraphics[width=0.48\textwidth]{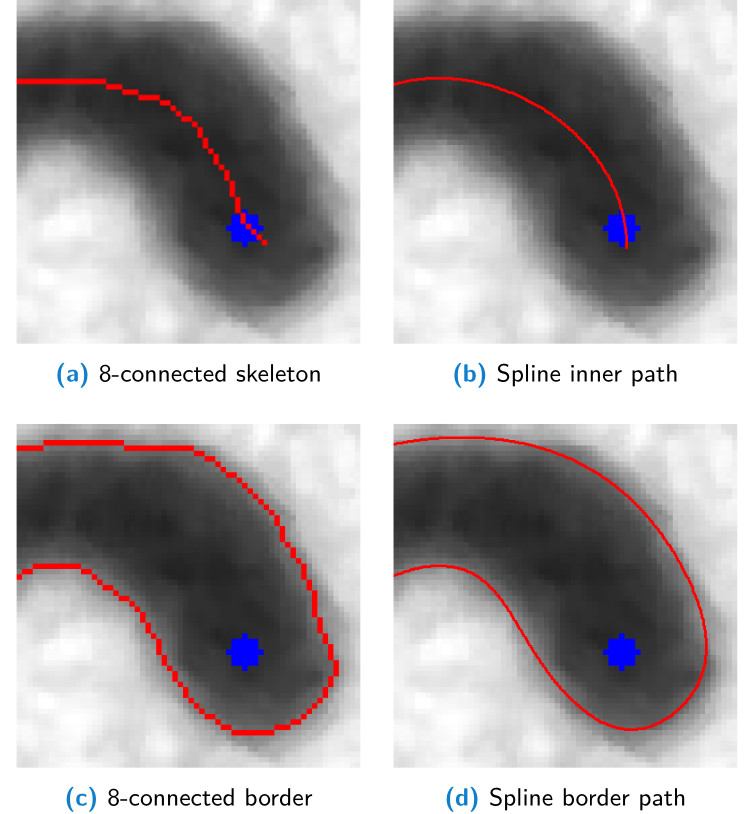}
    \caption{Example of 8-connected and spline-fitted paths for the inner (skeleton) and the border paths.}
    \label{fig:spline_fitting}
\end{figure}

Calculating path lengths directly from pixel paths introduces significant errors due to a fundamental limitation: the discrete pixel grid approximates continuous shapes as jagged stair-steps (discretization error), with diagonal paths particularly affected by connectivity rules (4-connectivity overestimates while 8-connectivity underestimates).
To minimize these effects, a smooth spline was fitted to the paths (Fig.~\ref{fig:spline_fitting}), allowing for estimation of the real Euclidean lengths of the stripes by numerically integrating positions sampled from it.

Smooth splines were generated through an interpolation process, sampling from the pixel paths using the {\it splprep} function from the SciPy library.
The degree of smoothing is determined by the coefficient $s$ which limits the total square error as shown in equation:

\begin{equation}
    \sum_j [(g(x_j) - y_j)]^2 \leq s
    \label{eqn:spline_smoothing}
\end{equation}

where $g(x_j)$ is the smoothed spline and $y_j$ are pixel samples from the path.
The value of $s$ was chosen as equal to the length of the path in pixels, to ensure that paths of different lengths were smoothed to a similar degree~\cite{scipy_smoothing_splines}.
Points on the spline were then sampled using {\it splev}, also from SciPy, with the number of samples being twice the number of pixels in the original path.
To generate the Euclidean length of each path, the spline was numerically integrated based on the sampled points.

Since the medial skeleton terminal does not necessarily reach the edge of the magnetic stripe, an additional length equal to the local half‑width (the Euclidean distance from the terminal to the nearest edge) was added to ensure accurate measurement of inner paths.
Additionally, for branches incident to a junction, the segment from the junction to the edge (the local half‑width at the junction) was subtracted from the length measurements.
Under this convention, a junction is treated as growing from the stripe’s border rather than its center.

\subsection{Curvature measurements}

Curvature measures how sharply a curve departs from a straight line at each point and is given by:

\begin{equation}
    \kappa = \frac{|x'y'' - y'x''|}{({x'}^2 + {y'}^2)^{\frac{3}{2}}}
    \label{eqn:curvature}
\end{equation}

We computed it for both the inner and border paths of the magnetic labyrinthine pattern using their parametrized splines.

The total curvature of each path, defined as the path integral $\int \kappa\,\mathrm{d}s$, was estimated numerically by sampling each spline at twice the original pixel resolution.
For each image, this total curvature was then divided by the corresponding path length, yielding the mean curvature of the inner and border trajectories.

\section{Results}

\subsection{Dark area}
\label{sec:dark_area}

\begin{figure}
    \centering
    \includegraphics[width=0.5\textwidth]{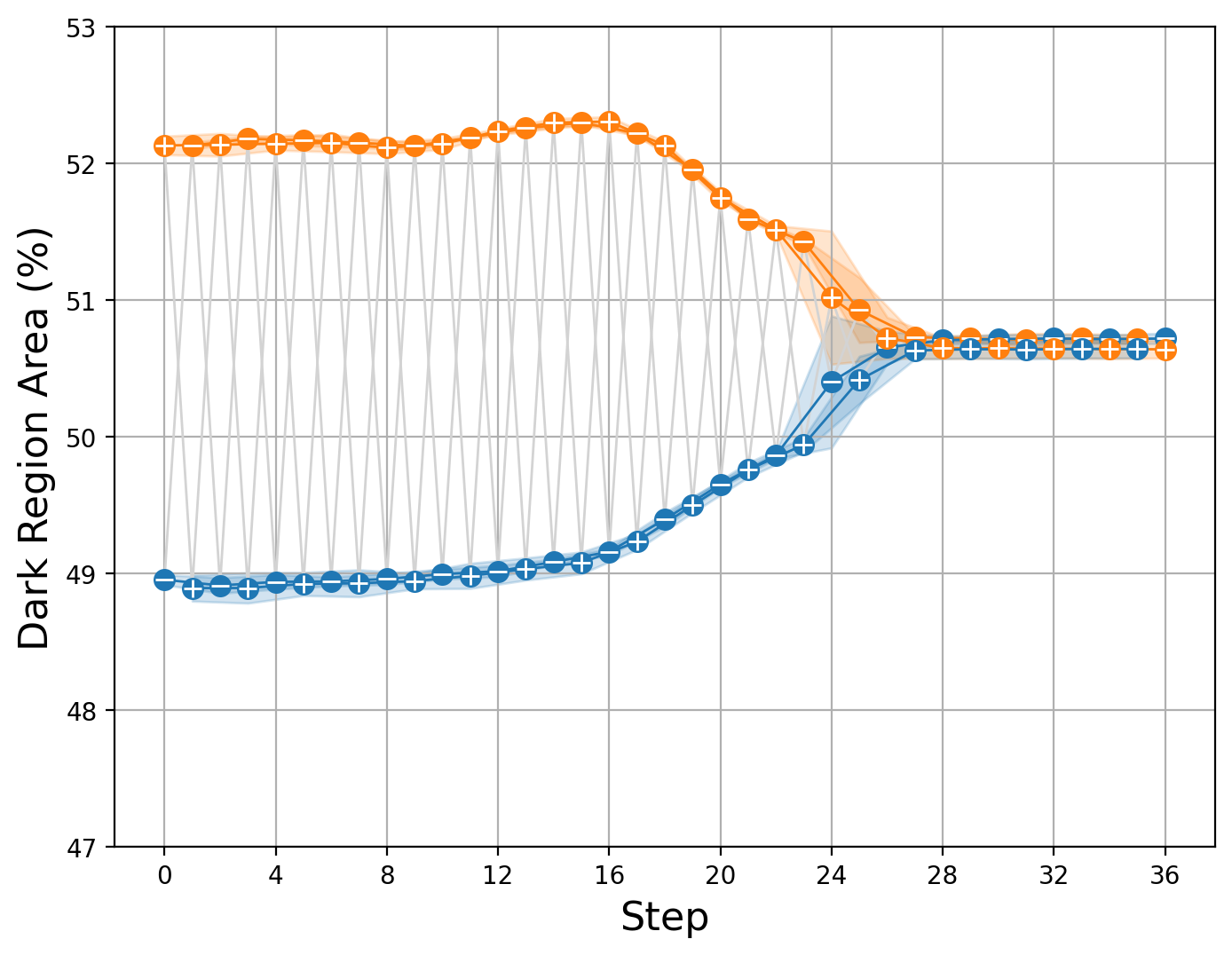}

    \caption{Dark region area as a function of step. The “+” markers indicate the positive process and the “–” markers the negative process. 
    Light-gray lines highlight the interleaved evolution in both processes, alternating between Type~A (orange) and Type~B (blue).}
    
    \label{fig:dark_region_area}
\end{figure}

From the U-Net segmented images, the area occupied by the dark region was estimated (Fig.~\ref{fig:dark_region_area}).
Measurements from the positive annealing protocol ($\sigma \in \{+\}$) are marked with “+”, while those from the negative protocol ($\sigma \in \{-\}$) are marked with “-”.
Results for both protocols were averaged over six trials each, with the standard deviation shown as a shaded region.
An interleaved pattern emerged in both the positive and negative sequences, depending on whether the measurement is Type~A (plotted in orange), where the applied field is aligned with the magnetization of the black region, or Type~B (plotted in blue), where it is aligned with the magnetization of the white region.

The area of a dark region---corresponding to a ferromagnetic domain---is governed by the interplay between the external magnetic field and competing spin interactions. In both quench and annealing protocols, the applied field plays a central role in shaping domain morphology through the Zeeman coupling, which energetically favors alignment of magnetic moments with the field. Another key contribution in thin-film systems is the magnetic anisotropy energy, which tends to align magnetic moments perpendicular to the film plane. The emergence of stripe patterns in such thin films reflects a competition between interactions at different length scales. Short-range exchange favors uniform spin alignment and thus large domains, while long-range dipole–dipole interactions penalize net out-of-plane magnetization, promoting alternating up- and down-magnetized regions. This frustration stabilizes a modulated stripe phase, with a characteristic width set by the balance between exchange, which favors broader domains, and dipolar interactions, which favor finer alternation.

During imaging, the magnetic field is zero, and thus an equal distribution of dark and bright areas is expected. Meanwhile, the annealing protocol involves a magnetic field quench and the majority of magnetic moments align with the field direction before the quench. This is particularly pronounced in the early stages, where fields above the saturation field are applied, driving the YIG film into a fully-polarized state. Due to the memory effect of the system, an imbalance between the dark and bright domain areas is observed. For instance, in Type~A measurements, a larger dark domain area is observed compared to Type~B.

In the later stages of the annealing protocol, the quenching magnetic field is smaller than the saturation field.
As a result, the system no longer resets to a fully polarized state. Instead, it can evolve into other labyrinthine patterns that are energetically close to its current configuration.
This facilitates the emergence of configurations with approximately 50$\%$ dark region area. Consequently, as the protocol progresses, both Type~A and Type~B measurements exhibit a convergence toward an equal area fraction.
In practice, the resulted segmentations have a slight bias towards dark regions.

\subsection{Defect count}
\label{sec:defect_count}

\begin{figure}
    \centering
    \includegraphics[width=0.5\textwidth]{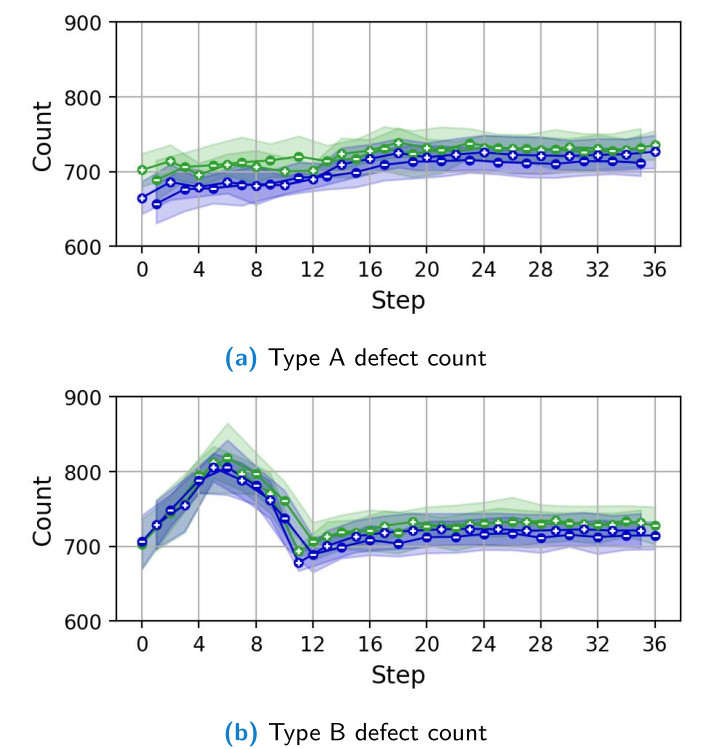}

    \caption{
    Number of defects averaged over six trials. Green illustrate the count of junctions, and blue show the count of terminals.
    }
    \label{fig:number_of_defects}
\end{figure}

Fig.~\ref{fig:number_of_defects} shows the evolution of defect counts during the annealing procedure. In Type~B measurements, the quenched state begins with a higher number of defects, which gradually decreases and stabilizes at a lower value as annealing progresses, consistent with Ref.\cite{okubo2024}. In contrast, Type A measurements start with a slightly lower defect count in the quenched state, which stabilizes at a higher value in the annealed state, eventually converging toward the defect count of Type~B. Although the step dependence differs between Type~A and Type~B, junctions and terminals within the same sequence display similar step-dependent trends. This behavior reflects the topological nature of these defects: junctions and terminals cannot be created or annihilated individually through smooth stripe deformations, but only through pair creation or annihilation~\cite{shimizu2023}.

Annealing is generally expected to increase coherence and reduce the number of defects. 
At first glance, the results in Fig. \ref{fig:number_of_defects} seem contradictory. 
Yet, in our measurements, the defect count does not monotonically decrease during the protocol.
This apparent inconsistency is resolved by considering the change in stripe period reported in Ref. \cite{shimizu2023}. 
During annealing, the stripe period decreases by roughly 20\% between steps 12 and 24. 
Since defects are tied to stripe structures, a shorter period naturally supports a higher defect density. 
Indeed, for Type~A measurements, the defect count normalized by the stripe period decreases~\cite{shimizu2023}. Since the defect count in Type~B measurements in the annealed state is larger than that in Type A, the normalized defect count in Type B likewise decreases.

Comparison of Type~A and Type~B measurements shows that before step 12 the defect count of Type~B is substantially higher, while beyond this step the difference between the two becomes negligible.
The qualitative differences in defect counts arise from distinct dominant dynamics at different stages: in the early steps, local defect dynamics prevail, while in the later steps the variation in stripe period becomes the primary factor, as previously reported~\cite{shimizu2023}.

In the early stage, the strong quench field resets the system to a fully polarized state at each step. 
The rapid field quench then nucleates local stripe seeds, which evolve under zero field toward a configuration where dark and bright regions ideally occupy equal areas. 
In Type~B measurements, where the dark-region fraction is smaller (Fig.~\ref{fig:dark_region_area}), the system compensates by generating additional stripe branches through the pair creation of junctions and terminals. 
As a result, the defect count exceeds that observed in the dark-area-dominated Type~A measurements.

In Type~B measurements, the defect count peaks near step 6. During the initial stages of demagnetization, successive steps generate additional defects, causing a steady increase. Between steps 6 and 12, however, the count decreases due to pair annihilation of topological defects. The creation of defects helps drive the system toward a balanced fraction of dark and light regions but simultaneously disrupts structural coherence. As demagnetization progresses, these defects are gradually annihilated, restoring coherence and giving rise to the observed peak around step 6.

Beyond step 12, the lack of a significant difference in defect counts between Type A and Type B measurements suggests that the spatial distribution and correlations of defects converge to similar states in both measurements. 
This occurs even though the distinct areas of dark and bright regions are maintained, as shown in Fig. \ref{fig:dark_region_area}. 
At this point, the evolution of the labyrinthine patterns is governed by a reduction in the stripe period, not by the local dynamics of topological defects. 
To achieve patterns with equal dark and bright areas, the labyrinthine stripes change their local period, leading to an increase in the total number of defects within a fixed area.

\subsection{Inner length between defects}

\begin{figure}
    \centering
    \includegraphics[width=0.5\textwidth]{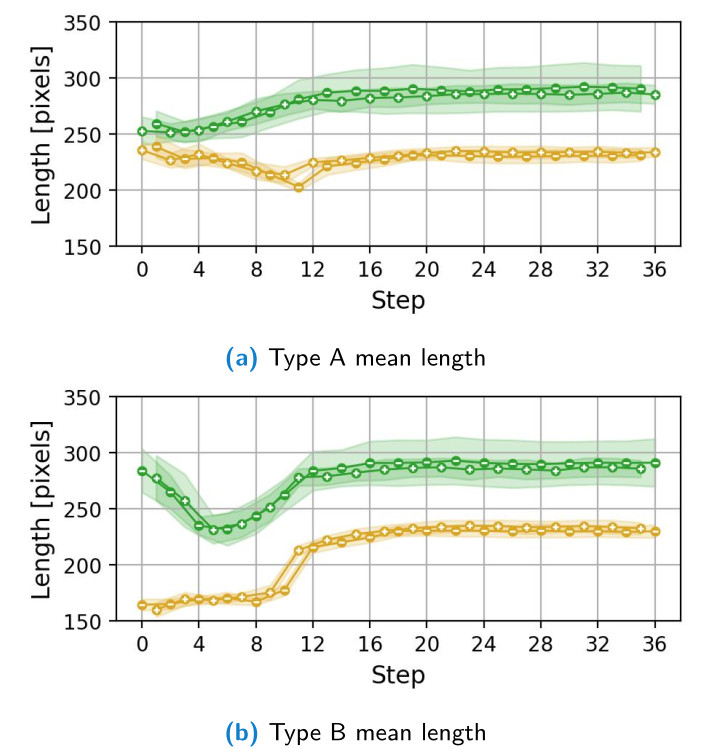}

    \caption{
    Average inner length measurements between topological defects, calculated across six trials.
    Green segments: Length between adjacent junctions.
    Yellow segments: Length between adjacent junctions and terminals.
    }
    \label{fig:mean_defect_distance}
\end{figure}

The mean inner axis length between defects is shown separately for junction–junction and junction–terminal pairs in Fig.~\ref{fig:mean_defect_distance}.
Each mean-length measurement exhibits a distinct trend across the annealing protocol for both Type~A and Type~B images.
Junction–junction pairs consistently display longer average lengths than junction–terminal pairs.
While junction-terminal pairs can undergo pair creation and annihilation, junction-junction and terminal-terminal pairs cannot directly change the overall defect count due to the topological properties. Considering that the system tends to evolve toward a more coherent structure by reducing defects, junction-terminal pairs attract each other more strongly than the other pairs. This difference in pair interactions is reflected in the shorter mean length observed for junction–terminal pairs.

In Type~B images, the junction–junction curve dips around steps 5–6, which appears to mirror the higher number of topological defects seen in Fig.~\ref{fig:number_of_defects}.
A larger number of junctions divides the stripe pattern into shorter segments, possibly explaining this dip.
However, the Type~A data and the Type~B junction–terminal measurements do not show a clear connection between length between defects and defect count.

The first factor to account for the discrepancy is that the distances between defects are calculated along the stripes. While the Euclidean distance between randomly distributed points decreases as the number of points increases, the distance along a stripe can increase if the stripe becomes more compactly folded, even with the same number of points. Indeed, as shown in Fig.~\ref{fig:magnetic_stripe_pattern}, in the later-stage annealed state, the stripes exhibit more frequent bending. The overall increasing trend, particularly in the lengths between junctions, can be attributed to this feature.

The second factor is the correlation between defects. In Ref.~\cite{shimizu2023}, it was reported that strong correlations develop between junctions and terminals, and around step 12, where the system undergoes a transition from the quenched to the annealed state, the annihilation of junction-terminal pairs contributes to an increase in the coherence of the system. Since short junction–terminal pairs are preferentially annihilated, the mean length of the remaining pairs can increase. This trend is especially pronounced in Type~B measurements for junction–terminal pairs.

\subsection{Curvature}

Defects discussed above serve as an indicator of geometrical inhomogeneity in the system and of deviations from the thermal equilibrium state. For stripe patterns in particular, their dynamics are crucial for systematically understanding the demagnetization dynamics from a topological perspective. On the other hand, for the relationship between the configuration of magnetic moments and the energy, the curvature provides a more direct and convenient quantity. As noted earlier, stripe formation arises from the competition between short-range exchange interactions and long-range dipole–dipole coupling. While exchange interactions favor uniform spin alignment, they also energetically prefer straight domain walls. Consequently, regions where black–white domain boundaries bend sharply carry a higher energy cost: the larger the local curvature, the greater the energetic penalty.

\begin{figure}
    \centering
    \includegraphics[width=0.48\textwidth]{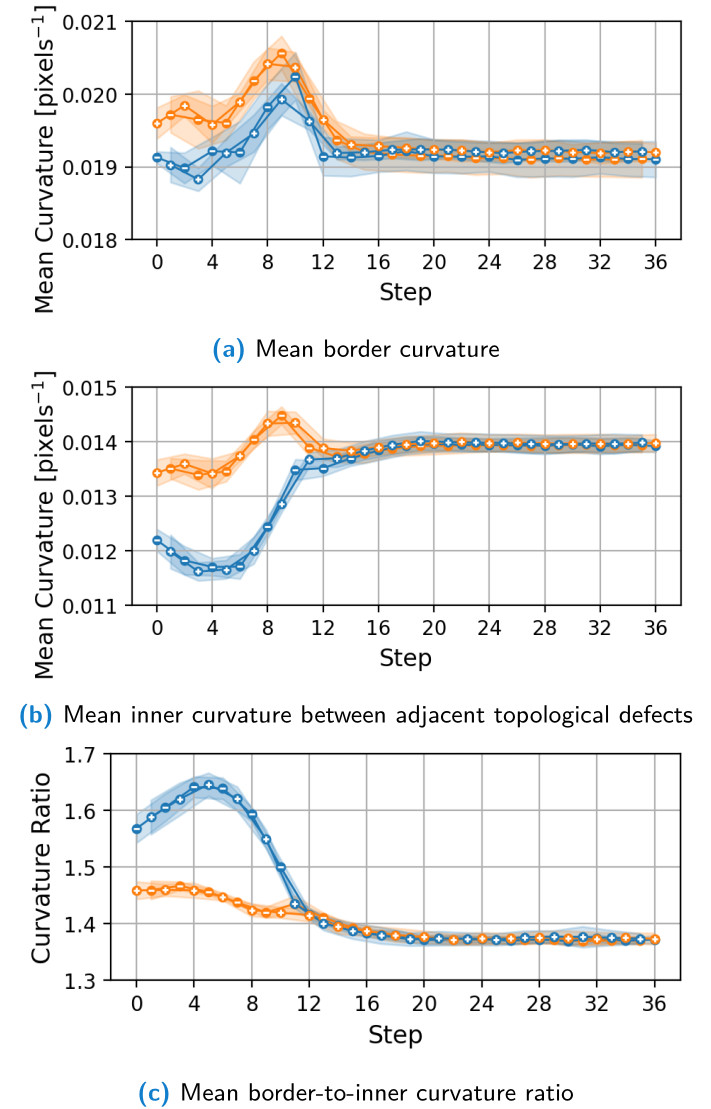}
    \caption{Curvature measurements averaged across six trials.
    Orange is Type~A measurements, blue is Type~B.}
    \label{fig:curvature_measurements}
\end{figure}

Fig.~\ref{fig:curvature_measurements}(a) displays the border curvature of the labyrinthine pattern. 
Both Type~A and Type~B images show similar curvature patterns, though Type~A has slightly higher curvature in the quenched state. 
For both measurements, the curvature peaks around steps 9-11, coinciding with the transition from the quenched to the annealed state.
The peak in curvature around step 10 may be related to the defect pair annihilation discussed in Section~\ref{sec:defect_count}.
Pair annihilation of defects requires substantial local deformation of the stripe pattern, which can produce sharp spatial modulations in the vicinity; consequently, the mean curvature may increase. After the annihilation process, however, the system locally relaxes to a smoother pattern with fewer defects, leading to a subsequent decrease in the curvature.

Fig.~\ref{fig:curvature_measurements}(b) presents the mean inner curvatures of paths between adjacent topological defects. 
Type~B images display lower inner curvature in the quenched state, likely reflecting the higher density of topological defects and shorter mean inter-defect distances at this stage. 
Because junctions redirect the stripes, a greater share of directional changes occurs at the junctions themselves, and stripe segments between defects exhibit less internal curvature.
Shorter segments also fit more easily into the magnetic pattern, aligning with the observed reduction in within-segment curvature. 
In contrast, Type~A images show more uniform mean inner curvature, consistent with their smaller variations in stripe length and defect count.

Fig.~\ref{fig:curvature_measurements}(c) presents the ratio between the border and inner curvature measurements.
Since changes in direction of the magnetic stripe is caused by both bending of the stripes and due to topological defects, this ratio can give a qualitative sense to with how much of the directional changes is due to the presence of the junctions and terminals.
Type~B measurements in particular, show a high curvature ratio during the initial steps, mirroring the high number of topological defects present in it.

\subsection{Total length}

\begin{figure}
    \centering
    \includegraphics[width=0.48\textwidth]{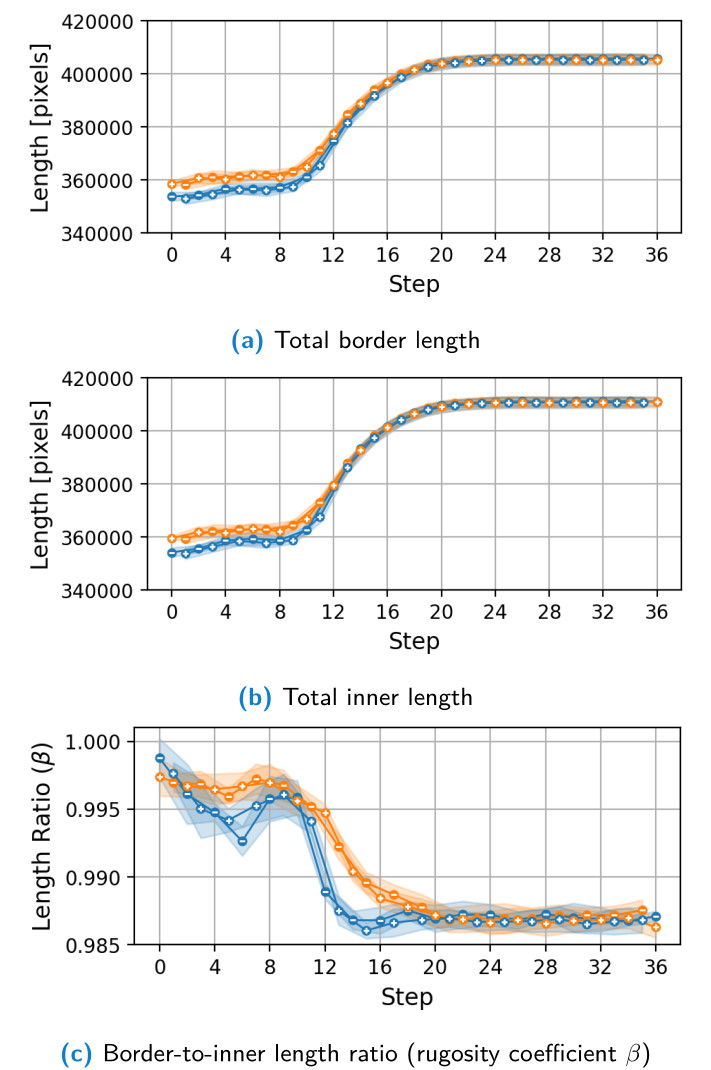}
    \caption{Length measurements averaged across six trials. Orange is Type~A measurements, blue is Type~B.}
    \label{fig:length_measurements}
\end{figure}

Fig.~\ref{fig:length_measurements} (a)-(b) illustrates the total inner and border lengths of the magnetic stripe pattern images.
The graphs show a lower average length at the beginning of the annealing protocol (in the quenched state), followed by stabilization at a higher average length in the later steps (in the annealed state).
This increase in total length throughout the annealing protocol can be attributed to a more efficient packaging of stripes in the annealed state due to its higher spatial order as well as the reduction of the stripe period~\cite{shimizu2023}.

Noticeably, the total lengths for Type~A and Type~B images are remarkably similar, despite high variation in geometric (lengths and curvatures between defects) and topological (junction and terminal counts) properties previously shown.
Type~A images are shown to have slight longer total lengths during the quenched state, which can be associated to the higher area occupied by the black stripes.

The ratio of border length to inner length provides a measure of magnetic stripe pattern border rugosity  (coefficient $\beta = L_{border} / L_{inner}$).
Fig.~\ref{fig:length_measurements} (c) showcases this coefficient during the annealing protocol.
During the initial steps in the quenched state a higher rugosity is observed. 
During the later steps, the rugosity decreases as the labyrinthine pattern approaches a more parallel configuration.

\section{Conclusion}

In this work, we presented a comprehensive methodology for the quantitative characterization of labyrinthine magnetic stripe patterns. Our approach integrates a robust U-Net deep learning model, trained with synthetically augmented data including additive white Gaussian noise and Simplex noise, to achieve precise segmentation of the intricate stripe patterns in Bi:YIG films. This robust segmentation enabled accurate characterization of boundaries and internal structures of the magnetic domains, even in the presence of real-world experimental artifacts.

Building upon this segmentation, a geometric analysis pipeline enabled the extraction of precise local measurements, including stripe length, curvature and the lengths of segments between topological defects. This analysis focuses on local features of the stripe propagation, allowing for characterizing important structural details which can be obscured in traditional global analysis methods.

Our analysis quantified the evolution of the labyrinthine patterns during a magnetic field annealing protocol, demonstrating the transition from a spatially disordered “quenched” state to a more parallel and compact “annealed” state. 
By distinguishing between “Type~A” and “Type~B” measurements based on magnetic field polarity, we uncovered intertwined behaviors across various properties, including the dark region area, defect count, defect-to-defect lengths, and curvature. 
These findings offer novel insights into the complex interplay of geometric and topological factors governing the propagation and evolution of magnetic stripe paths.

The ability to precisely measure local properties and track their evolution under external stimuli opens new avenues for understanding the fundamental physical mechanisms underlying pattern formation, as well as for the potential control and manipulation of such fascinating structures. 
Future work could expand on the physical interpretations of the observed intertwined behaviors and extending this methodology to other such complex pattern-forming systems.

\section*{Acknowledgements}
This study was partially funded by the São Paulo Research Foundation (FAPESP), Brazil (grant numbers 2024\allowbreak10263-3 and 2025/03683-9), and by the National Council for Scientific and Technological Development (CNPq), Brazil (grant 300724/2025-0).
The work of BSS at the University of Virginia was partially funded by the NSF under grant DMR \#2016909.
The work of GWC at the University of Virginia was partially supported by the US Department of Energy Basic Energy Sciences under Contract No. DE-SC0020330.
The Article Processing Charge (APC) of this research publication was funded by the Coordenação de Aperfeiçoamento de Pessoal de Nível Superior – CAPES (ROR identifier: 00x0ma614). For the purposes of open access, the authors have applied the Creative Commons CC BY license to any accepted version of the article.

\appendix

\section{Noise details}
\label{sec:appendix_A}

\begin{table}[h]
    \caption{Parameters used for synthetic degradations.}
    \label{tab:noise_parameters}
    \centering
    \begin{tabularx}{\columnwidth}{lXX}
    \toprule
    \textbf{Noise Type} & \textbf{Parameters} & \textbf{Value} \\
    \midrule   
    Simplex & Threshold & [60, 150] \\
           & Scale & [200, 500] \\
           & Octaves & [4, 6] \\
           & Persistence & [0.4, 0.6] \\
           & Lacunarity & [2, 3] \\
    \midrule
    AWGN & $\sigma$ & [0, 120] \\
    \midrule
    Photometric & $\alpha$ & [0.8, 1] \\
            & $\gamma$ & [0.4, 1] \\
    \bottomrule
    \end{tabularx}
\end{table}

\begin{table*}[ht]
    \caption{Visual comparison of segmentations using Otsu's method and the U-Net architecture with various data augmentation schemes: no synthetic degradations, Gaussian noise and photometric transformation, only Simplex noise, and all combined synthetic degradations.}
    \label{tab:detailed_comparison}
    \centering
    \setlength{\tabcolsep}{3pt}
    \begin{tabularx}{\textwidth}{
        @{}
        C{0.079\textwidth}
        *{6}{C{0.142\textwidth}}
        @{}
    }
        \toprule
        \textbf{Condition}
        & \textbf{Original}
        & \textbf{Otsu}
        & \textbf{U-Net (no degradations)}
        & \textbf{U-Net (AWGN + photometric)}
        & \textbf{U-Net (Simplex)}
        & \textbf{U-Net (All augmentations)} \\
        \midrule
        Clear
            & \includegraphics[width=\linewidth,valign=c]{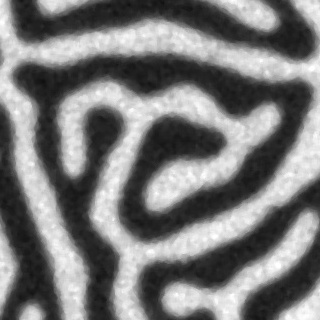}
            & \includegraphics[width=\linewidth,valign=c]{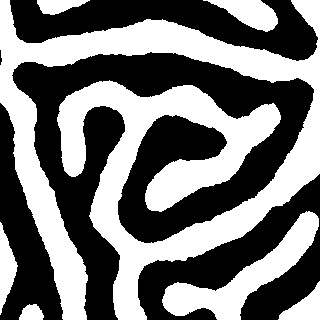}
            & \includegraphics[width=\linewidth,valign=c]{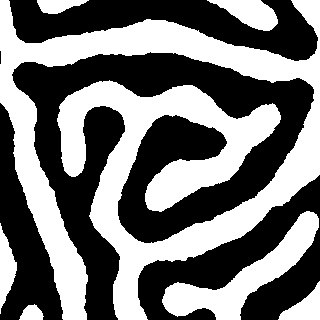}
            & \includegraphics[width=\linewidth,valign=c]{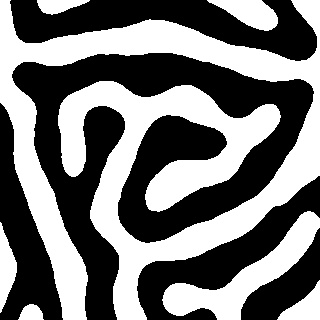}
            & \includegraphics[width=\linewidth,valign=c]{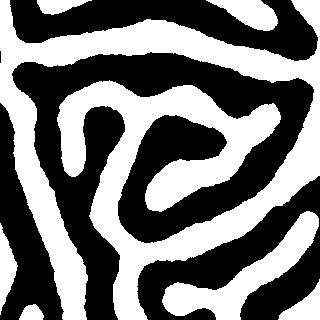}
            & \includegraphics[width=\linewidth,valign=c]{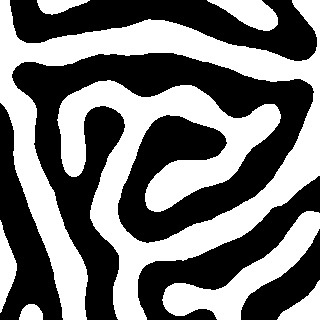} \\
        Noisy
            & \includegraphics[width=\linewidth,valign=c]{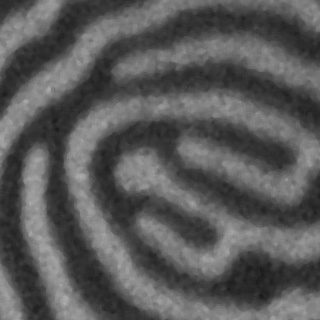}
            & \includegraphics[width=\linewidth,valign=c]{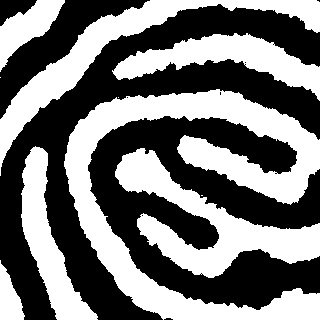}
            & \includegraphics[width=\linewidth,valign=c]{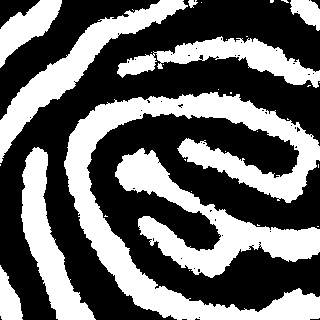}
            & \includegraphics[width=\linewidth,valign=c]{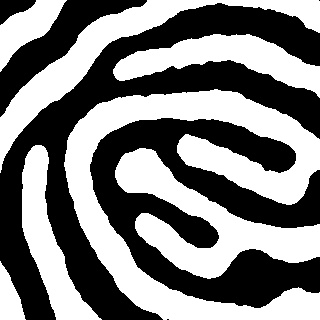}
            & \includegraphics[width=\linewidth,valign=c]{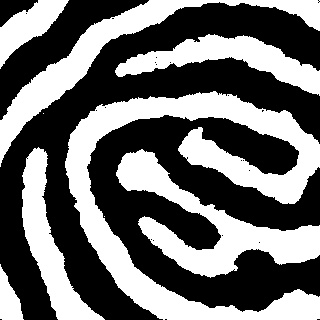}
            & \includegraphics[width=\linewidth,valign=c]{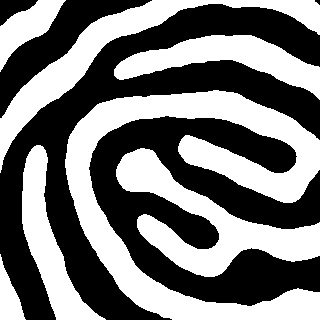} \\
        Occluded
            & \includegraphics[width=\linewidth,valign=c]{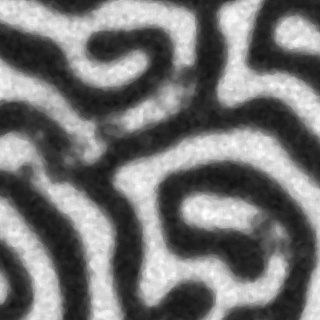}
            & \includegraphics[width=\linewidth,valign=c]{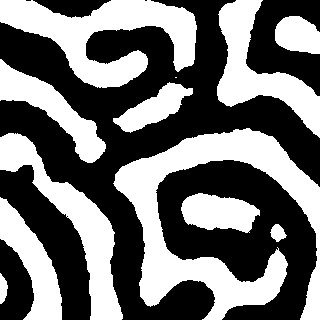}
            & \includegraphics[width=\linewidth,valign=c]{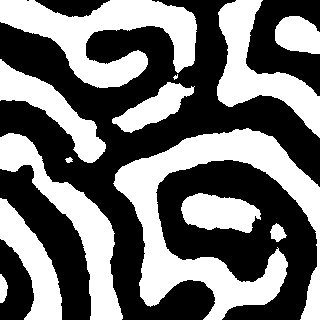}
            & \includegraphics[width=\linewidth,valign=c]{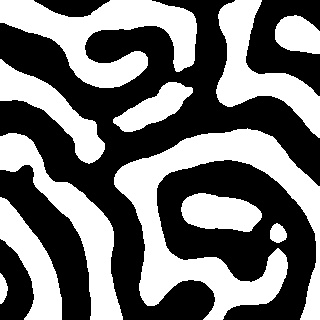}
            & \includegraphics[width=\linewidth,valign=c]{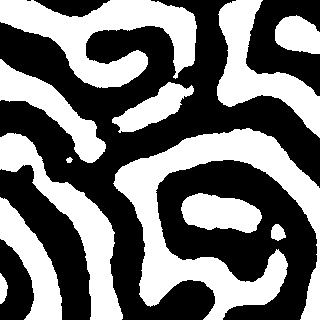}
            & \includegraphics[width=\linewidth,valign=c]{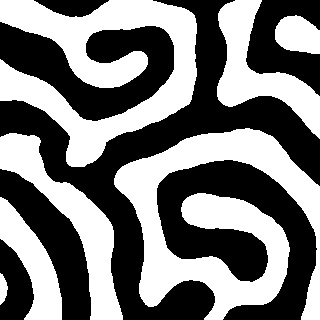} \\
        Noisy and Occluded
            & \includegraphics[width=\linewidth,valign=c]{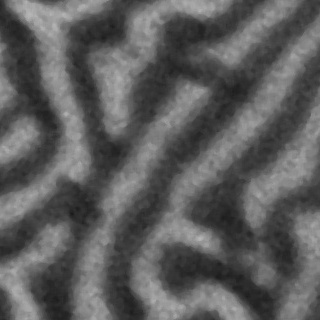}
            & \includegraphics[width=\linewidth,valign=c]{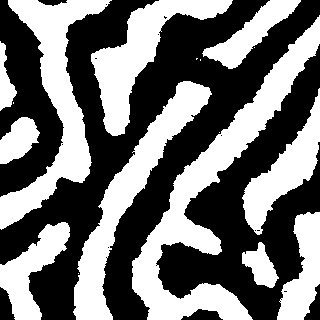}
            & \includegraphics[width=\linewidth,valign=c]{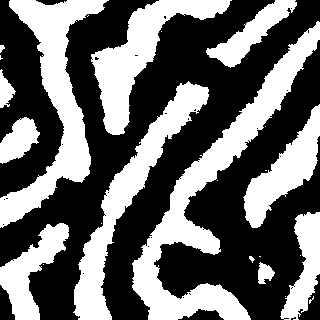}
            & \includegraphics[width=\linewidth,valign=c]{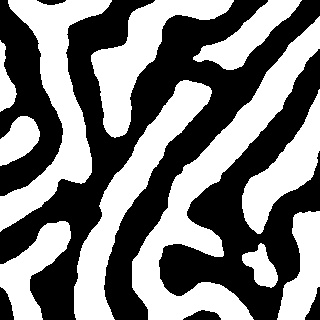}
            & \includegraphics[width=\linewidth,valign=c]{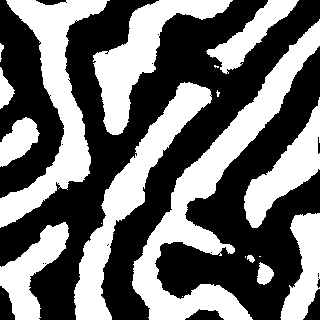}
            & \includegraphics[width=\linewidth,valign=c]{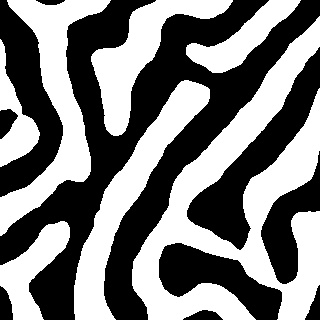} \\
        \bottomrule
    \end{tabularx}
\end{table*}

In order to account for varying degrees of degradation, synthetic noise was generated with parameters uniformly sampled from predefined ranges (Table~\ref{tab:noise_parameters}). These ranges were selected through visual inspection and kept intentionally broad to ensure that visually plausible degradations are well covered across the sampled parameter space.

For the Simplex noise occlusions, the parameters \textit{scale}, \textit{octaves}, \textit{persistence}, and \textit{lacunarity} control the texture of the generated patterns, while the \textit{threshold} parameter determines the size of the occluded regions.
Specific visual patterns are not directly associated with individual parameter values, instead resulting from the nonlinear interaction between them.
We individually varied each parameter and checked whether visually plausible occlusion degradations would still occur under the parameters joint behavior.

Simplex noise was added first, sampling values from the scale, octaves, persistence and lacunarity parameter ranges.
It was scaled to the $[0, 255]$ range and  then applied to the patch according to:

\begin{equation}  patch_{\mathrm{occluded}}
  = \min\!\left( \frac{simplex\_noise \cdot 255}{threshold},\; patch \right)
\label{eqn:simplex_application}
\end{equation}

This process was repeated twice to generate overlapping Simplex noises with different parameter values, improving robustness.

Additive white Gaussian noise was then applied once, followed by the photometric transformation.
Lastly, the image patches and the segmented patches are rescaled to the $[0, 1]$ range prior to training and inference.

Table~\ref{tab:detailed_comparison} presents a visual comparison on the effect of different augmentation noise configurations and the resulting segmentation. To verify robustness, the table compares segmentation results across four scenarios: clear regions, noisy regions, occluded regions, and regions with both noise and occlusion.

Without any data augmentation, U-Net's segmentations visually resembled Otsu thresholding, containing both irregular boundaries and occlusions.
Training the model with both AWGN and photometric transformation resulted in smoother and better-defined borders between light and dark regions, although issues with occlusion degradations persisted.
When trained exclusively with Simplex noise, the segmentation was similar to Otsu's method in clear regions, with moderate improvements in boundary smoothness and occlusion filtering.
When all augmentations were used during training, segmentation borders remained smooth and well-defined, and filtering of occluded regions further improved.
These results suggest that Simplex noise alone does not adequately model true occlusions found in real magnetic labyrinthine images,  leading to limited generalization to real occlusions.

\section{Validation metrics}
\label{sec:appendix_B}

Validation performance metrics are presented in table~\ref{tab:validation}.
An ablation study on the impact of different synthetic degradations in training is provided.
Using all AWGN, photometric transformation and Simplex noise yields the best overall balance between overlap, boundary distance, and length deviations.

\begin{figure}
    \centering
    \includegraphics[width=0.48\textwidth]{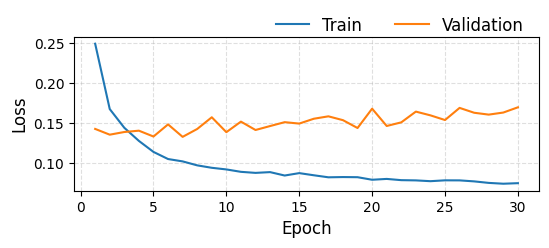}
    \caption{Validation loss curve.}
    \label{fig:loss_curve}
\end{figure}

\begin{figure}
    \centering
    \includegraphics[width=0.48\textwidth]{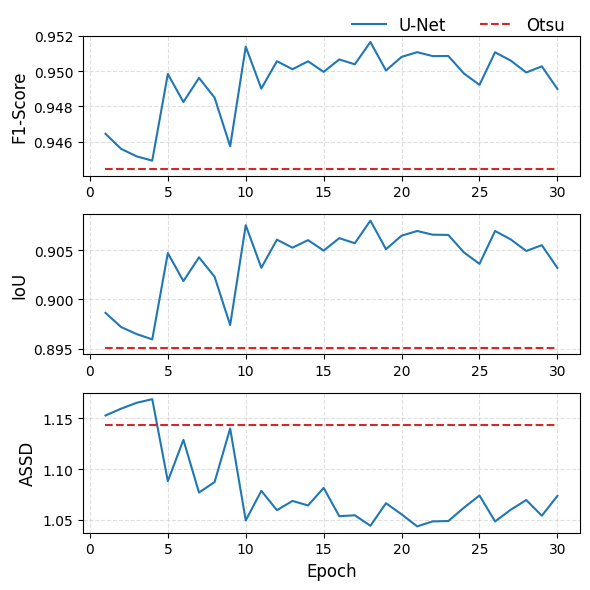}
    \caption{Validation performance curves.}
    \label{fig:val_metrics}
\end{figure}

Fig.~\ref{fig:loss_curve} showcase the validation loss curve of the U-Net model using all augmentations, while Fig.~\ref{fig:val_metrics} showcase the validation performance metrics throughout training.

\begin{table}[h]
    \centering
    \caption{Validation performance metrics.}
    \label{tab:validation}
    \setlength{\tabcolsep}{3pt} 
    \begin{tabularx}{\columnwidth}{@{} l @{} *{5}{>{\centering\arraybackslash}X @{} } }
        \toprule
        Model &
        \makebox[0pt][c]{F1 ($\uparrow$)} &
        \makebox[0pt][c]{IoU ($\uparrow$)} &
        \makebox[0pt][c]{ASSD ($\downarrow$)} &
        \makebox[0pt][c]{$\Delta$Inner} &
        \makebox[0pt][c]{$\Delta$Border} \\
        \midrule
        Otsu & 0.944 & 0.895 & 1.14 & 3.33\% & 3.43\% \\
        SegFormer-B3 & 0.951 & 0.907 & 1.06 & -0.76\% & -0.93\% \\
        U-Net (all augs.) & \textbf{0.952} & \textbf{0.908} & \textbf{1.04} & -0.57\% & \mbox{\textbf{-0.10\%}} \\
        \unetvar{No aug.} & 0.940 & 0.888 & 1.24 & 3.33\% & 4.79\% \\
        \unetvar{AWGN+photo.} & 0.947 & 0.900 & 1.12 & \textbf{0.01\%} & 0.59\% \\
        \unetvar{Simplex} & 0.943 & 0.893 & 1.19 & 1.19\% & 2.29\% \\
        \bottomrule
    \end{tabularx}
\end{table}

\printcredits

\bibliographystyle{cas-model2-names}

\bibliography{refs}



\end{document}